\documentclass[journal,comsoc]{IEEEtran}
\usepackage[T1]{fontenc}
\usepackage[nocompress]{cite}
\usepackage[pdftex]{graphicx}
\usepackage{microtype}
\usepackage{booktabs}
\usepackage{xcolor}
\usepackage{todonotes}

\usepackage{amsmath}
\usepackage[cmintegrals]{newtxmath}
\usepackage{bm}

\usepackage{eqparbox}

\newcommand{\augusto}[1]{{\color{red}A: #1}}
\newcommand{\diego}[1]{{\color{blue}D: #1}}
\newcommand{\milton}[1]{{\color{green}M: #1}}

\usepackage{array}

\usepackage[caption=false,font=normalsize,labelfont=sf,textfont=sf]{subfig}
\usepackage{dblfloatfix}
\usepackage[hyphens]{url}
\usepackage{hyperref}
\usepackage[hyphenbreaks]{breakurl}

\usepackage[shortcuts]{extdash}

\begin{document}
%
\title{Permissionless Refereed Tournaments}
%
%
%

\author{Diego~Nehab\quad and\quad Augusto Teixeira%
\IEEEcompsocitemizethanks{\IEEEcompsocthanksitem Authors are affiliated to IMPA and Cartesi, e-mail: diego.nehab@cartesi.io, augusto.teixeira@cartesi.io}}

\maketitle

\begin{abstract}
Scalability problems in programmable blockchains have created a strong demand for secure methods that move the bulk of computation outside the blockchain.
One of the preferred solutions to this problem involves off-chain computers that compete interactively to prove to the limited blockchain that theirs is the correct result of a given intensive computation.
Each off-chain computer spends effort linear on the cost of the computation, while the blockchain adjudicates disputes spending only logarithmic effort.

However, this effort is multiplied by the number of competitors, rendering disputes that involve a significant number of parties impractical and susceptible to Sybil attacks.
In this paper, we propose a practical dispute resolution algorithm by which a single honest competitor can win disputes while spending effort linear on the cost of the computation, but only logarithmic on the number of dishonest competitors.
This algorithm is a novel, stronger primitive for building permissionless fraud-proof protocols, which doesn't rely on complex economic incentives to be enforced.
\end{abstract}


%
\IEEEpeerreviewmaketitle

\section{Introduction}

%
%
%
%
\IEEEPARstart{T}{raditional} consensus protocols~\cite{pease1980reaching,lamport1982byzantine} like \emph{Paxos}~\cite{lamport1998part} and \emph{Raft}~\cite{ongaro2014search} presuppose a known set of players.
For some types of applications, however, this permissioned setup raises trust concerns since players would have undue authority over those not allowed to participate in the game.
On the other hand, \emph{permissionless} protocols (in which players are not known beforehand and may dynamically join and leave the game) are susceptible to \emph{Sybil attacks}~\cite{douceur2002sybil} (a malicious entity may create multiple false identities and shift the consensus in their favor).

The \emph{Nakamoto protocol}~\cite{nakamoto2008bitcoin,garay2015bitcoin}, used in blockchain networks like Bitcoin, introduced a novel solution for Sybil attacks that makes launching an attack exponentially expensive (by cleverly using ``proofs of work''~\cite{dwork1992pricing} and hash chains~\cite{lamport1981password}).
Later, in programmable blockchain networks like Ethereum~\cite{wood2022ethereum}, the protocol was extended to support smart contracts: ``Turing-complete'' applications executed as part of the blockchain's consensus, but submitted by anyone, and with which anyone can engage.

Nevertheless, the protocol is not without its flaws.
Although safe (it can resist a large number of players trying to attack it) and decentralized (it can run without any trust dependencies on a small group of players), it is rather inefficient.
Naive attempts to make networks faster end up sacrificing safety and/or decentralization, in what has become known as the \emph{scalability trilemma}~\cite{Sharding-FAQs}.
For example, one can tune the protocol by choosing different reference computers, in a way that only supercomputers would be able to run the protocol (i.e., validate the network).
This increases scalability but harms safety and decentralization, since fewer players would be able to participate.

The scalability issue, however, is more insidious than just making sure old laptops can validate the network; it is related to the global nature of the consensus.
Usually, players are only interested in a few applications of a blockchain, rather than all of them.
This is similar to traditional web applications, where users are only interested in a tiny subset of the entire internet.
Unlike traditional web applications, however, every node in a blockchain must validate all its applications (i.e., global consensus), rather than just what they are interested in.
Naively distributing the load across nodes dilutes validation power, making the whole network inherit the safety of the weakest link.
Although adding more computers makes a network safer by increasing validation power, it doesn't increase execution power.
There's a limit on what a network can execute, dictated by what a single reference computer can process.

This will never scale to the volumes seen in web applications: no single computer can run the entirety of the internet.
The consequence is highly gentrified networks, wherein users have to outbid each other for the right to use a slice of the blockchain, in ever growing fees.
This is undesirable.
After all, ``permissionless'' only to the wealthy isn't really permissionless.

Fortunately, there's light at the end of the tunnel.
Rollups~\cite{Rollups} are a promising solution to the blockchain scalability problem.
In fact, Ethereum has its own rollup-centric roadmap~\cite{rollup-roadmap}.
In its essence, a rollup instance is an off-chain computational layer (i.e., a layer-2) on top of an existing network (i.e., the base-layer or the layer-1).
The goal is to move applications from the base-layer to layer-2s, shifting as much of state and computation as possible from the global consensus into independent local consensuses.

The key challenge is to ensure the results of executing a layer-2 can be efficiently validated by the base-layer, which is ultimately responsible for enacting any global side effects (e.g., monetary transfers, final public record, etc.).
This type of problem, in which a computationally powerful prover---or set of provers---must convince a computationally limited verifier of the validity of a computation, has a long history in computer science~\cite{Feige1997,gennaro2010non,goldwasser2015delegating}.
A popular technique of proving the correctness of an execution is refereed games~\cite{Feige1997,CanettiRivaEtAl2013}, which has been adopted by several layer-2 projects~\cite{TeutschEtAl2017,Arbitrum2018,TeixeiraNehab2018}\footnote{Rollups that use this fraud-proof approach are called \emph{optimistic rollups}.}.
This technique gives the base-layer the ability to efficiently referee a dispute between two layer-2 players that disagree on the result of a computation over layer-2 state.

The general idea is so simple it fits within a single paragraph (but see Section~\ref{sec:two-party} for more details).
Layer-2 computations are run in deterministic, self-contained machines that maintain a Merkle tree~\cite{Merkle1979} over their entire state.
The initial state of the machine contains all the information necessary to perform the desired computation and it can be summarized by the root hash of the Merkle tree.
Likewise, the final result is summarized by the root hash after the computation has been performed: the final state hash.
In case of a dispute over two candidate final hashes, the base-layer interactively guides both parties in a binary search to find the step before which the state hashes match, but after which they disagree.
At this point, the base-layer queries one of the participants for the information needed for the simulation of the single offending step while progressively updating the state hash.
If the resulting hash matches the hash that was claimed by the queried participant, they win the dispute.
Otherwise, they lose.
The dispute protocol therefore guarantees that an honest claim will prevail.
(See~\cite{CanettiRivaEtAl2013, TeixeiraNehab2018} for further details.)

Consider a computation that takes $t$~steps and $s$~memory.
Obtaining the final state hash in layer\=/2 takes $O(t + k s)$ effort, where~$k$ is a security parameter (e.g., the hash size).
Settling a dispute between two participants takes $O(k \log t + k \log s)$ effort in layer\=/1 and $O(t + k s \log t)$ in layer\=/2.
Crucially, the protocol creates virtually no layer\=/2 overhead when participants agree, and little overhead even when they engage in a dispute.
As an indication of the computational scalability that is achievable, one of these projects actively encourages disputable layer\=/2 computations to be run under the Linux operating system~\cite{TeixeiraNehab2018}.

Refereed games have at least two disadvantages.
First, the dispute resolution protocol is interactive.
This means that layer\=/2 transactions cannot be finalized until every participant with a stake in the computation has had the chance to dispute its result
(although liquidity providers can mitigate the delayed finality).
Second, disputes between large numbers of participants are still problematic.
This second problem is the focus of our work; permissionless refereed tournaments are the primitive over which layer-2s want to build their fraud-proof protocols, otherwise the number of players allowed to participate in disputes has to be restricted.
This raises trust concerns since players would have undue authority over those not allowed to participate in the game,
which is counterproductive to blockchains since it would make disputes permissioned.

There are two existing algorithms for multi-party disputes~\cite{CanettiRivaEtAl2013}.
In the first approach, each participant must simultaneously engage in a dispute against every participant with which it disagrees.
This quickly becomes impractical due to limits in the computing power of participants.
The alternative is to run as many sequential disputes as there are claims.
In that case, a dispute involving $p$~parties takes, per party, $O\big(p(k \log t + k \log s)\big)$ computation from layer\=/1 and $O\big(p (t + k s \log t )\big)$ from layer\=/2.
When $p$ is large, this leads to potentially unacceptable delays.
Current implementations try to side-step the problem by restricting the layer\=/2 participants allowed to engage in disputes to a small committee~\cite{Cartesi_Rollups}, or require claimants to post collaterals that must grow with the total value staked in layer\=/2~\cite{Arbitrum2018}.
Both approaches harm decentralization.

In this paper, we show how to effectively eliminate this limitation.
Our contributions are as follows:
\begin{itemize}
\item We present a novel algorithm that enables an honest party to win any dispute with $O\big((\log p)(k \log t + k \log s)\big)$ effort in layer\=/1 and $O\big(t k\log s + (\log p) (k \log t+t+k \log s)\big)$ effort in layer\=/2;
\item We then propose a variation of the algorithm that runs in $O\big(\tfrac{t}{n} k \log s + (\log p) (k\log t+t+k \log s)\big)$ layer\=/2 time. Constant~$n$ can be chosen large enough to mitigate the effect of the $k \log s$ multiplier to $\tfrac{t}{n}$ in the original solution;
\item Finally, we showcase a prototype implementation of the algorithm using Cartesi's RISC-V emulator~\cite{TeixeiraNehab2018}. Our implementation can settle multi-party disputes between a large number of participants performing general computations under the Linux operating system.
\end{itemize}

In summary, our solution gives honest competitors an exponential computational advantage over dishonest competitors.
Moreover, since each party must pay for their own share of layer\=/1 cost, the advantage is also financial.
Honest parties are effectively protected against Sybil attacks in which a dishonest party creates a large number of colluding pseudo-identities.
We believe these contributions will drive a new generation of effective layer\=/2 scalability solutions for blockchains.

\section{Related work}
\label{sec:related-work}

The problem of delegating intensive computations has been approached using a variety of techniques that offer different tradeoffs in terms of efficiency and theoretical guarantees.

In a typical setup, this problem involves a computationally limited referee that needs the result of an computation so intensive it does not have the power to perform itself.
Instead, this referee relies on one or many powerful but untrusted third-parties that must participate in a protocol to reach consensus.

As mentioned in the introduction, this problem is relevant for blockchain applications because their expensive consensus mechanism naturally limits the amount of computations that can be performed on the blockchain itself.
Meanwhile, it is often the case that only a small group of clients are interested in a particular computation and they would be willing work off-chain to reach local consensus.
In principle, this would free them from the limitations imposed by the consensus mechanism, allowing for a much higher performance.
But at the same time, if these parties happen to disagree on the computation results, they would like to be able to settle their disputes using the full security and decentralization offered by the layer\=/1.

There are many proposed solutions to this problem and one way in which they differ is the security assumptions they make.
Some offer computationally-sound proofs~\cite{micali2000computationally}, or rely on the absence of hash collisions~\cite{ben2018scalable,CanettiRivaEtAl2011}, or on assumptions from number theory (like the difficulty of calculating discrete logarithms~\cite{bunz2018bulletproofs}).

Another difference between these solutions is the method by which the referee verifies results.
Some methods require that each claim comes accompanied with a \emph{validity proof}.
Often, these are cryptographic proofs that can be verified for correctness in a fraction of the original cost of performing the computation \cite{groth2018updatable}.
Unfortunately, the creation of such proofs is several orders of magnitude more computationally demanding than the original computation.

Other protocols optimistically allow a given party to claim the computation results, but offer a time-window for honest parties to challenge such claims.
These solutions differ widely in the type of challenge mechanisms they implement.

In the simplest approach, the challenges happen in a single step, but require the blockchain itself to perform the computation in full~\cite{Optimism_Rollups}.
Despite their simplicity, these methods limit the scalability gains to what the layer\=/1 is able to verify directly.

More elaborate protocols guarantee the validity of claims through an interactive verification protocol, where many queries and answers are exchanged between the parties involved \cite{Feige1997,CanettiRivaEtAl2013}.
The payoff for these interactions is that these solutions provide much greater scalability, as discussed below.

See table~\ref{t:complexity} for a brief comparison of the computational complexities of different approaches, which are described in more detail below.

\begin{table*}[t!]
  \caption{Comparison between algorithms for delegated computation.}
  \label{t:complexity}
  \centering
  \setlength{\tabcolsep}{3pt}
  \begin{tabular}{cccc}
    \toprule
    Algorithm & Layer-2 commitment & Layer-2 dispute & Layer-1 verification \\
    \cmidrule(l{2pt}r{2pt}){1-3}
    \cmidrule(l{2pt}r{2pt}){4-4}
    Optimism \cite{Optimism_Rollups} & $t + ks$ & $t + ks$ & $t$ \\
    Canetti et al. \cite{CanettiRivaEtAl2013} & $t + ks$ & $p(t + ks\log t)$ & $ p (k \log t+ k \log s)$ \\
    Ours (single-stage) & $t k \log s$ & $(\log p)(k\log t + t + k\log s)$ & $ (\log p)  (k \log t + k \log s)$ \\
    Ours (multi-stage) & $ \tfrac{t}{n} k \log s $ & $(\log p) (k\log t+t+k \log s)$ & $ (\log p)  (k \log t + k \log s)$ \\
    \bottomrule
    \multicolumn{4}{l}{$t$ is the number of steps in the computation. $s$ is the amount of memory used in the computation.} \\
    \multicolumn{4}{l}{$k$ is a security parameter. $n$ is a constant large enough to mitigate the effect of the $k \log s$ multiplier to $\tfrac{t}{n}$. }
  \end{tabular}
\end{table*}

\subsection{Interactive verification}
\label{sec:Interactive_Verification}

The most prominent approaches to offload intensive computations from the blockchain consensus layer involve~\emph{interactive verification}.

The first of these algorithms was introduced in~\cite{Feige1997}.
It recruits two computationally powerful parties to perform a computation and then submit their claims to the referee.
In case of disputes, the referee can adjudicate in favor of the honest party after a logarithmic number of interactions.
A major advantage of this idea is that it offers an exponential increase in efficiency for the referee, while maintaining the linearity of the execution cost for the recruited parties.
The original algorithm was later significantly improved by the use of collision-resistant hash functions~\cite{CanettiRivaEtAl2011}, which allowed for the elimination of the private communication channels between the referee and the participants

With fraud proofs, there is no proof that a given party has submitted the correct result of the computation.
Instead, the algorithm enables an honest party to \emph{disprove} a competing claim by a dishonest party.
For this, however, the honest party must engage in a dispute.
Even in its improved form, therefore, interactive verification does not scale well when the number of participants grow.
This is a very serious issue in un-permissioned blockchains, where any malicious actor can generate a variety of pseudo-identities to overwhelm honest parties, which must refute every false claim.
(This is often called a \emph{Sybil attack}.)

Nevertheless, this algorithm has been implemented on blockchain solutions by various projects \cite{TeutschEtAl2017,Arbitrum2018,TeixeiraNehab2018}.
Each of these projects has proposed a different solution to the Sybil vulnerability above.
Such mitigations include simple restrictions on the number of participants that can take part in the computation~\cite{Cartesi_Rollups}, security deposits that can be used to penalize dishonest behavior~\cite{Arbitrum2018} and very elaborate crypto-economic protocols~\cite{TeutschEtAl2017}.

In this work we present a non-economic solution to this problem that scales exponentially well with the number of participants.
Moreover, if our algorithms is coupled with economic incentives, we can guarantee that the honest party can recover a reward proportional to the number of dishonest parties $p$, while spending only $\log p$ during the dispute.

\section{Permissionless Refereed Tournaments}

To better motivate our solution to the multi-party dispute resolution problem, we will describe the steps we followed in its conception.
To that end, will discuss two alternative solutions and why they both fail.
For completeness, we start with a formal description of the two-party dispute resolution algorithm of~\mbox{Canetti et al.}~\cite{CanettiRivaEtAl2013} using the notation of Teixeira and Nehab~\cite{TeixeiraNehab2018}.

\subsection{Preliminary definitions}
\label{sec:preliminary-definitions}

Let~$S$ denote the entire state of a deterministic, self-contained machine, including all its memory, drives, and the value of all its registers.
(We assume the state is divided into words, that each word is denoted by its address~$a$ as $S[a]$, and that $|S|=s$.)
Let each instruction in the machine architecture take constant time to execute and let~$\textsc{step}$ be a function that advances the state by executing the next instruction, so that $S_{i+1} = \textsc{step}(S_{i})$.
In this context, any desired computation can be specified by an initial state~$S_0$.
Let the computation comprise the evolution of states via the $\textsc{step}$ function until a final state~$S_t$ is reached that is a fixed-point of~$\textsc{step}$, i.e., $S_t = \textsc{step}(S_{t})$.

We can associate a \emph{state hash} $M$ to each machine state~$S$ by defining~$M$ as the root hash of a Merkle tree~\cite{Merkle1979} built over the words in the machine state.
For the sake of completeness, we describe the construction and key properties of Merkle trees.
Given a cryptographic hash function~$H$ (for example, SHA-3~\cite{Dworkin2015}), the security parameter~$k$ measures the size of the hash as well as the cost-per-byte of evaluating the hash function.
Let~\mbox{$H = \textsc{hash}(D)$} denote the hash associated to data~$D$.
Assume the number of words in the state is a power of two (pad it with zeros otherwise) and let \mbox{$d = \log s$}.
To create the tree, we start by applying the hash function to each word individually~\mbox{$H^d_i = \textsc{hash}(S[i])$}.
The resulting hashes of each pair of consecutive words are then concatenated and hashed again.
The process is repeated, so that at each iteration~$\ell$ we have $H^\ell_i = \textsc{hash}(H^{\ell+1}_{2i} H^{\ell+1}_{2i+1})$.
A binary tree is thus built bottom-up, where each node is labeled by a hash, until the hash~$H^{0}_0$ that labels the tree root is found.
This is the value of~$M = \textsc{state-hash}(S)$.

It is possible to efficiently accept or reject claims concerning the value~$w$ of a word at address~$a$ in the state, given only its state hash~$M$ and a small amount of auxiliary data.
The idea is to progressively reconstruct all hashes $\{P^d,P^{d-1},\ldots,P^1,P^0\}$ in the path between the hash of the word in question and the Merkle tree root.
The first hash in the sequence is obtained from the claimed word itself as~$P^d=H^d_a=\textsc{hash}(w)$.
The value for the word is accepted if the final hash matches the state hash, i.e., if $P_0=M$.
The intermediate hashes can be produced from the siblings $\{Q^{d-1},\ldots,Q^{0}\}$ of all nodes in the path.
This is the auxiliary data needed to prove a claim.
The process is embodied by function~$\textsc{up}$, with

\begin{small}
\begin{align}
\textsc{up}(a, \{\}, H) &= H \\
\textsc{up}(a, \{Q^{d-1},\ldots,Q^{0}\}, H) &=
\textsc{up}(\lfloor \tfrac{a}{2} \rfloor, \{Q^{d-2},\ldots,Q^{0}\}, H'),
\end{align}
\end{small}
where
\begin{small}
\begin{align}
H' &= \begin{cases}
\textsc{hash}(H Q^{d-1}), & \text{$a$ even} \\
\textsc{hash}(Q^{d-1}\! H), & \text{$a$ odd}.
\end{cases}
\end{align}
\end{small}

To accept $\{Q^{d-1},\ldots,Q^{0}\}$ as proof that the word at address~$a$ has value $w$ in a state with hash~$M$, simply check if
\begin{align}
\textsc{up}\big(a, \{Q^{d-1},\ldots,Q^0\}, \textsc{hash}(w)\big) = M.
\end{align}
The proof size and verification effort are both~$O(k\log s)$.
The idea can be trivially extended to accept or reject claims concerning the hash that label any internal node in the tree.

Remarkably, it is also possible to construct a counterpart function~$\textsc{step-hash}$ to $\textsc{step}$ that operates entirely on hashes.
$\textsc{step-hash}$ requires only a small amount of auxiliary data in the form of an \emph{access log} that effectively serves as a proof of state transition.
Formally, given an access log~$L_i$, $\textsc{step-hash}$ must either reject the log as invalid, or return the new state hash~$M_{i+1} = \textsc{step-hash}(M_i, L_i)$ such that $S_{i+1} = \textsc{step}(S_i)$, with $M_i = \textsc{state-hash}(S_i)$ and $M_{i+1} = \textsc{state-hash}(S_{i+1})$.

The \textsc{step} function can be simulated by a constant number of reads and writes to the state, controlled by the logic that implements the machine's instruction set architecture.
The \textsc{step-hash} function can use the exact same logic as~$\textsc{step}$, so long as \textsc{step-hash} receives a log of all state accesses \textsc{step} performs in the process of advancing the state.

Consider a step simulated with~$\ell$ state accesses.
The~$\ell$ log entries allow \textsc{step-hash} to update the state hash through all necessary modifications $M_i = M^0_i, M^1_i,\ldots$ until it reaches $M^k_i=M_{i+1}$.
For each entry~$j\in\{1,\ldots,\ell\}$, the access log contains
\begin{enumerate}
\item the type of access $t_j\in\{\mathit{read}, \mathit{write}\}$;
\item the address $a_j$ of the access;
\item the word value~$r_j$ at $a_j$ in $M_i^{j-1}$;
\item the siblings hashes $\{Q^{d-1},\ldots,Q_0\}_j$;
\end{enumerate}
While running the simulation, the logic in $\textsc{step-hash}$ decides that the $j$th access happens at address $a_j'$ and must either read the corresponding word, or write a new value $w_j'$ in its place.
In case of reads, the function
\begin{enumerate}
\item checks $t_j = \mathit{read}$;
\item checks $a_j = a_j'$;
\item checks \smash{$\textsc{up}\big(a_j, \{Q^{d-1},\ldots,Q^0\}_j, \textsc{hash}(r_j)\big) = M_i^{j-1}$};
\item maintains the state hash~$M_i^j = M_i^{j-1}$.
\end{enumerate}
It can now use the value of $r_j$ to advance its logic, confident it was indeed the value at $a_j'$ in the state.
In case of writes, it
\begin{enumerate}
\item checks $t_j = \mathit{write}$;
\item checks $a_j = a_j'$;
\item checks \smash{$\textsc{up}\big(a_j, \{Q^{d-1},\ldots,Q^0\}_j, \textsc{hash}(r_j)\big) = M_i^{j-1}$};
\item updates \smash{$M_i^j = \textsc{up}\big(a_j, \{Q^{d-1},\ldots,Q^0\}_j, \textsc{hash}(w_j')\big)$}.
\end{enumerate}
In summary, to accept~$L_i$ as proof that~$M_{i+1}$ is the hash of the state that follows a state with hash~$M_i$, simply check if
\begin{align}
M_{i+1} = \textsc{step-hash}(M_i, L_i)
\end{align}

The access log size and verification effort are also~$O(k\log s)$.
Moreover, the $\textsc{step-hash}$ function can only be fooled into accepting a fake access log that advances the state incorrectly in the presence of hash collisions.
By the assumed resistance of the hash function, designing such collisions is beyond the reach of any adversary.

\subsection{Two-party dispute resolution}
\label{sec:two-party}

With definitions above, we can formally describe the two-party dispute resolution protocol of~\mbox{Canetti et al.}~\cite{CanettiRivaEtAl2013}.
Consider a machine with initial state hash $M_0$, to be run until it halts or for at most $t$ steps, and two parties~$A$ and~$B$.
The referee obtains from~$A$ and~$B$ the commitments~$F_A$ and~$F_B$ for the final state hash.
If they match, there is no dispute.
Otherwise, the referee guides $A$ and~$B$ in a binary search.
The result is a step in the computation before which $A$ and~$B$ agree the state hash is~$N$, but after which~$A$ claims it is~$I_A$ and $B$ disagrees.
The referee then asks~$A$ for the access log~$L$ that allegedly advances the state hash from~$N$ to~$I_A$.
Invoking its own implementation of the $\textsc{step-hash}$ function on access log~$L$, the referee advances the state.
If the result is~$I_A$, then $A$ wins the dispute.
Otherwise, $B$ wins.

From the referee's point of view, the protocol takes effort $O(k\log t + k\log s)$.
The first term corresponds to the binary search.
The final term corresponds to the evaluation of the $\textsc{step-hash}$ function based on the access log.
From the point of view of the disputing parties, the effort is~$O(t + k s\log t)$.
When the binary search interval is~\mbox{$[u, v]$}, players keep a copy of state $M_u$ before advancing to $M_m$, where $m=\lfloor \tfrac{u+v}{2}\rfloor$.
This saves the effort of restarting from~$M_0$ whenever the interval is narrowed to $[u, m]$.
Therefore, instead of spending effort~$O(t\log t)$ advancing the state, the expenditure is only~$O(t + s \log t)$.
Computing the state hash takes time~$O(k s)$, and this is repeated once per iteration in the binary search.
Altogether, the cost is $O(t + s\log t + k s\log t) = O(t + k s\log t)$.

\subsection{Linear multi-party dispute resolution}
\label{sec:linear-multi-party}

When there are~$p$ parties claiming a result for a computation, \mbox{Canetti et al.}~\cite{CanettiRivaEtAl2013} suggest a straight-forward modification to the protocol described above.
Let~$P$, $|P|=p$, be the list of parties involved in the dispute.
The referee first obtains a list~$F$ with the final states claimed by each party.
If they are all the same, there is no dispute.
Otherwise, the referee enters into an iterative elimination process.
At each iteration, it uses a binary search to identify a step before which \emph{every remaining party} agrees the state hash is~$N$, but after which the list of state hashes~$I$ proposed by the remaining parties has at least two distinct entries.
The referee then obtains a list~$L$ with the access logs that each party claims advances the state to the state they propose in~$I$.
Using the list of access logs~$L$, the referee then uses its~$\textsc{step-hash}$ function to advance the state according to each log.
The parties for which the resulting state hash does not match the corresponding entry in~$I$ are eliminated from $P$ (and the corresponding entries eliminated from $F$).
The iterations continue until only parties that agree on the final state hash remain.
Since at least one party is eliminated at each iteration, the algorithm is guaranteed to terminate after at most~$p$ iterations.

It should be clear that, within each elimination iteration, all parties operate in the same way as they would in the two-party version of the algorithm.
Therefore, the honest party spends effort~$O\big(p(t + k s\log t)\big)$ in layer\=/2 and $O\big(p(k\log t + k\log s)\big)$ in layer\=/1.

\subsection{Failed attempts at improvement}
\label{sec:failed-improved-multi-party}

Computer scientists are naturally drawn towards the obvious approach to multi-party disputes in sub-linear time in~$p$: setting up a tournament bracket between all parties.
The naive approach starts with the referee, once again, collecting a commitment from each party concerning the final state hash.
When the final state hashes are all different, each match in the tournament can be decided by the two-party dispute resolution algorithm of section~\ref{sec:two-party}.
The honest party will win each of its matches, and eventually the entire tournament, after playing a number of matches logarithmic in~$p$.

This simple idea fails when multiple parties (some of which are dishonest) claim to agree on the final state hash.
Eventually, two such parties will be matched in a dispute.
This presents a serious problem because a dishonest party has the power to lose disputes on purpose, even when originally committed to the correct final state hash.
This means that an honest party cannot, in general, allow itself to be represented by another party in a dispute.
Since it is impossible to eliminate either party in a match when they agree on the final hash, they must be merged into \emph{teams}, and matches cease to be two-party disputes to become disputes between these teams.
At this point, we are forced to fall back to the linear multi-party dispute of section~\ref{sec:linear-multi-party}.
Otherwise, the binary search cannot proceed because members in a team may disagree on the intermediate state hash.
In other words, even intermediate hashes would be amenable to dispute.
In the worst case, the recursion allows a colluding team of dishonest parties to force an honest party to produce an access log for each step in the computation, which means the referee might as well perform the computation itself.

Blockchain specialists are instead drawn to a cryptoeconomic solution: put in place economic incentives that make disputes unaffordable (or at least irrational) to dishonest parties.
The leading proposal~\cite{Arbitrum2018} requires parties that wish to engage in a dispute to first deposit a collateral.
The winner of the dispute receives back its collateral and half the collateral from the losing party.
The remaining half is \emph{burned}.
A coordinated attack becomes impractical if the repeated collateral losses outweigh the benefit accrued from the delay in finality.
The problem here is that this solution requires the collateral to be proportional in value to whatever is at stake in the result of the computation.
If the collateral is fixed, then it is possible to envision scenarios in which delaying finality indefinitely is the rational choice.
On the other hand, if the collateral is proportional, it is possible to envision scenarios in which either the stakes must be limited, or only wealthy parties will ever be able to dispute a computation (i.e., decentralization is reduced).

Although we were unable to find a cryptoeconomic solution without these significant drawbacks, the tournament approach, as it turns out, \emph{can} be saved.

\subsection{Logarithmic multi-party dispute resolution}

The tournament idea in section~\ref{sec:failed-improved-multi-party} fails because parties are free to lie during the binary search.
Note that there is no such possibility during the final step of the algorithm, where parties must present an access log that allegedly advances the state between two known state hashes.
This is because the access logs can be directly verified by the referee.
The key contribution in our logarithmic multi-party dispute resolution algorithm is a means for the referee to directly verify binary search claims.

Our solution is to require all parties to make stronger initial commitments.
Rather than simply committing to a final state hash, they commit to a \emph{computation hash}.
A computation hash is the root hash of a Merkle tree whose leaves are the state hashes of each step through which the machine goes while performing a given computation.
Computation hashes therefore lock the temporal evolution of a machine's state, where the snapshot of each machine state is itself locked by a state hash.
Assume a machine~$M_0$ is to be run until it halts or for at most $t$~steps.
For simplicity, let~$t$ be a power of~2.
The corresponding computation hash~\smash{$C_{M_0}$} is the root hash of a complete Merkle tree of depth $d=\log t$.
If there are fewer than~$t$ steps, the last state hash is repeated to complete the tree.

To see the idea in practice, consider the following algorithm.
The referee starts by collecting a list~$C$ with the computation hashes from each party in~$P$.
A computation hash is accepted only if accompanied by a valid proof that the first state hash matches~$M_0$.
The parties are then partitioned into teams that posted the same computation hash.
A tournament bracket is setup in which each match is between two such teams, say, $A$ committed to $C_A$ and~$B$ committed to~$C_B$.

A match starts with a binary search for the earliest state hash the teams disagree on.
Crucially, since all members of a team are committed to the same computation hash, there is no possibility of disagreement.
The end result of the binary search is an address~$a$ and three state hashes~$N$, $I_A$ and~$I_B$ such that $N$ is the state hash at address~$a-1$ in both computation hash Merkle trees with roots~$C_A$ and~$C_B$, and $I_A$ and $I_B$ are the state hashes corresponding to address~$a$ in $C_A$ and $C_B$, respectively.
The search proceeds down both computation hash Merkle trees.
The referee keeps, for the current tree level, the address~$a$ and the labels~$H_A$ and~$H_B$ of the nodes in both trees.
It

\begin{enumerate}
\item initializes $a\gets0$, $H_A\gets C_A$ and $H_B\gets C_B$;
\item obtains from \emph{any} member of team~$T$, $T\in\{A,B\}$, the child labels~$\{H_T^\ell, H_T^r\}$ such that~$\textsc{hash}(H_T^\ell H_T^r)=H_T$;
\item if a timeout elapses before labels~$\{H_T^\ell, H_T^r\}$ are received, eliminates the entire team~$T$;
\item if $H_A^\ell=H_B^\ell$, then $a\gets 2a+1$, $H_A\gets H_A^r$, and~$H_B\gets H_B^r$,
otherwise  $a\gets 2a$, $H_A\gets H_A^\ell$, and~$H_B\gets H_B^\ell$;
\item loops back to 2)~$d-1$ times, until leaves are reached;
\item obtains state hash~$I_T$, $T\in\{A,B\}$ of step~$a$, from \emph{any} member of team~$T$,
along with proof $\{Q^{d-1},\ldots,Q^{0}\}_T$ satisfying $\textsc{up}\big(a, \{Q^{d-1},\ldots,Q^{0}\}_T, I_T)=C_T$;
\item does the same for state hash~$N_T$ of step~$a-1$. Note that, by construction, $N_A = N_B = N$;
\item if a timeout elapses before $N_T$ and $I_T$ are received, eliminates entire team~$T$.
\end{enumerate}
After the binary search is thus completed, deciding the winning team is trivial.
The referee
\begin{enumerate}
\item obtains from \emph{any} member of team~$T$, $T\in\{A,B\}$, the access log $L_T$ satisfying~$\textsc{step-hash}(N_T, L_T)=I_T$;
\item if $L_T$ is accepted, then eliminates the other team and declares~$T$ as the winner, otherwise, if a timeout is elapsed before~$L_T$ is received, eliminates the entire team~$T$.
\end{enumerate}

From the perspective of the referee, each match requires effort~$O(k\log t + k\log s)$.
The honest party will engage in $\log p$ matches, so its layer\=/1 effort is $O\big((\log p)(k\log t + k\log s)\big)$.
Each party obtains the computation hash in~$O(t k\log s)$ effort.
For the $\log p$ matches, the cost is $O(k\log t + t + k\log s)$.
Here, $k\log t$ is spent in the binary search and $t+k\log s$ is spent to advance until the step of disagreement and then generate the access log.
Thus, the total layer\=/2 effort spent by each party is~$O\big(t k\log s + (\log p) (k \log t+t+k \log s)\big)$.

In theory, the algorithm just described solves the scalability problem in dispute resolution: the number of dishonest parties, no matter what strategy they follow, has at most a logarithmic impact in the computational effort spent by an honest party defending its honest claim.
In practice, the algorithm imposes a substantial impact in that it makes every step significantly more expensive than the execution of a single instruction (the $k\log s$ multiplier to $t$) and it requires parties to store a Merkle tree with one leaf for each step in the computation (which can become prohibitive for long computations).
Fortunately, it is possible to address these problems with a modification to the algorithm.

\subsection{Multi-stage logarithmic multi-party dispute resolution}
\label{sec:multi-stage}

The idea is to split tournaments into~$b$ stages.
In a stage~1 tournament, all teams are committed to a \emph{sparse computation hash} that covers the entire computation, but that only records state hashes every~$2^{g_1}$ steps.
Constant~$g_1$ is chosen large enough to reduce the overhead of both storing and computing the state hashes and their associated Merkle tree.
In a stage~2 tournament, all teams are committed to a sparse computation hash that covers the interval between two state hashes in the stage~1 sparse computation hash, and records state hashes every $2^{g_2}$~steps, with $g_2 < g_1$.
Finally, in a stage~$b$ tournament, teams are committed to a complete computation hash that locks every computation step in the interval between two state hashes in the stage~\mbox{$b-1$} sparse computation hash (i.e., $g_b = 0$).

The referee first receives a stage~1 sparse computation hash from each party.
After checking each party's proof that the first state hash matches the initial hash for the computation, the referee uses the computation hashes to partition the parties into teams.
The dispute starts as a stage~1 tournament bracket, formed by stage~1-matches between teams within which each party agrees on the same stage~1 sparse computation hash.
A stage~1-match starts with a binary search that identifies the pair of consecutive state hashes (and their corresponding step indices) that bracket the disagreement between the two teams.
There is no possibility of disagreement between team members during the binary search.
All parties in both teams are then required to post the stage~2 sparse computation hash covering this sub-interval.
Here, there is room for disagreement between team members concerning the value of intermediate state hashes.
Therefore, each stage~1 team is partitioned into (potentially) multiple stage~2 teams, each newly created team committed to a different stage~2 sparse computation hash.
The resulting stage~2 teams are matched in a stage~2 tournament bracket, formed by stage~2-matches, that replaces the stage~1-match.
This process is repeated until stage~$b$ tournaments are reached.
These can be settled directly by the single-stage algorithm described in the previous section.

The effort spent by each party building the stage~1 sparse computation hash is $O(\tfrac{t}{n} k \log s)$, where $n = 2^{g_1}$.
The idea is to select~$n$ large enough to significantly reduce the number of state hashes computed and prevent the $k \log s$ cost from dominating the computation time.
For stage~$i$, $i\in\{2,\ldots,b\}$, the cost is $O(2^{g_{i-1}-g_i} k\log s)$, and therefore does not depend on~$t$.
Nevertheless, the number of stages and the number of steps skipped in each stage can be chosen so the constant cost of producing the computation hash for each stage is suitably small.
The stage~$1$ binary search for the interval of disagreement takes effort~$O(k \log \tfrac{t}{n})$.
In total, the effort spent by a party defending its claim is~$O\big(\tfrac{t}{n} k \log s + (\log p) (k\log\! \tfrac{t}{n}+t+k \log s)\big)$.
The cost of other stages is dominated by the cost of stage~1.

\section{Prototype implementation}

The Cartesi Machine~\cite{TeixeiraNehab2018} is an emulator for the RISC-V ISA~\cite{WatermanAsanovic2017U,WatermanAsanovic2017P}.
RISC-V is an open ISA created in UC Berkley for implementation by native hardware.
It is now maintained by its own independent foundation with the backing of large corporations, such as Google, Samsung, and Tesla~\cite{Tilley2018}.
The platform is supported by a vibrant community of developers.
Their efforts have produced an extensive software infrastructure, most notably ports of the Linux operating system and the GNU toolchain, both of which are upstream.

Cartesi's emulator has been designed specifically to support the dispute resolution protocols of~\mbox{Canetti et al.}~\cite{CanettiRivaEtAl2013}.
As such, it has some unique properties.
First, it is self-contained: it can be run under completely specified initial conditions and isolated from any external sources of entropy.
Second, it is reproducible: starting the emulator with the same initial state, even in different hosts, always produces the same progression of states.
Finally, it includes all primitives described in section~\ref{sec:preliminary-definitions} through well-documented APIs for the C++ and Lua programming languages.
The machine can be initialized with a user-configurable amount of RAM and a number of input/output ``flash drives''.
It can then be run for any specified number of steps with a quick snapshot/rollback feature.
Moreover, the machine exposes its entire state in a linear 64-bit address space over which a Merkle tree is constructed.
It can generate Merkle-tree proofs for the value of any word in the state.
Finally, it can generate the access log for each execution step, and includes a function that accepts or rejects an access log as proof of transition between two state hashes.

Using the Cartesi Machine API, we built an asynchronous TCP/IP server that plays the role of referee.
The server is given the initial state hash of a Cartesi Machine about to perform a desired computation.
Clients representing the parties that wish to participate in potential disputes then connect to the server.
The server then guides all parties in the protocol described in section~\ref{sec:multi-stage}.
Clients answer requests for computation hashes covering specific step intervals and periodicities, for the labels of children of any node in the Merkle tree of a computation hash, and for access logs that correspond to specific steps.
Once the protocol completes, the server prints out the winning team and the associated final state hash.

In order to force disagreement between players and properly test the protocol in action, we need a way to create dishonest Cartesi Machines.
For obvious reasons, Cartesi Machines are honest.
To implement a dishonest machine, we encapsulate an honest one in a wrapper class that obeys the same API, but that can be programmed to cheat by modifying the state of the underlying honest instance, in any way, at any prescribed step.
Our dishonest machines can also modify the access log generated for any step, in an (ultimately futile) attempt to make it consistent with an invalid state transition.

\subsection{Simple experiment}
\label{s:simple_ex}

Our first experiment is based on one of the RISC-V unit tests~\cite{RISCV-tests} for the \texttt{addi} instruction (addition with immediate operand).
This test, with 248 steps, is about as complex as a typical smart contract that runs directly on top of the blockchain.
It runs in \emph{machine mode}, in a flat address space, and does not use any advanced feature of the RISC-V ISA.
This makes it particularly simple to create dishonest machines that cheat at different moments in its execution.
For example, one dishonest machine modifies the memory contents or the value of a register before or after a particular instruction is executed.
Another changes the instruction itself that is about to be executed.
By carefully choosing these interventions, the dishonest machine can obtain any desired result.
In total, we produced 279 different dishonest machines for this experiment.

\subsection{Complex experiment}

Our other experiment, in a fictitious energy market, is more involved and requires some elaboration.
The details of the energy market are not relevant.
(We do not presume to be specialists, or even knowledgeable, in this subject.)
Instead, our goal is to showcase the type and intensity of verifiable computation now accessible to blockchain applications: the experiment runs 10,248,000,888 instructions.
Just as important, it runs under Linux (with virtual-memory support), and employs a wide range of standard development tools in its implementation that were previously unavailable---even unthinkable---in blockchain applications.

Assume each household in the energy market is equipped with smart meters that measure, minute by minute, their electricity consumption.
There are two sources of energy: hydroelectric and natural gas.
The hydroelectric plant can satisfy only part of the peak demand; the remainder must be supplied by natural gas.
In order to penalise nonrenewable energy usage, the following pricing scheme was designed.
For each minute in the day, the households are partitioned into quartiles by consumption.
The energy supplied by natural gas is paid for by each quartile using weights that reward low consumption and penalize excessive consumption.
More specifically, quartile~$i$ pays~$\tfrac{i}{10}$ of the natural gas costs.

At the end of each day, each smart meter consolidates its minute-by-minute consumption into a CSV file, signs it, and makes the file and signature available to an aggregator.
The aggregator compresses the CSV files and signatures together into a single data blob (say, using the Brotli compressor~\cite{Brotli2016}), builds a Merkle tree root hash for the blob, signs the root hash, and posts both the root hash and the signature to the blockchain.
The blockchain only accepts the entry if the signature matches the certificate it has for the aggregator.
The aggregator then makes the data blob publicly available off-chain to all households.

The computation under potential dispute in this scenario is the amount owed by each household.
This computation happens inside a Cartesi Machine configured with 2 input flash drives and 1 output drive.
Input drive~1 is initialized with a root file system that contains an embedded Linux distribution, including an assortment of executables such as the Tar, Brotli, an OpenSSL utilities.
In addition, it carries a Python~3 interpreter loaded with all its standard libraries, most notably the SQLite support.
The root file system comes preloaded with the certificates for all households.
Input drive~2 contains the compressed data blob distributed by the aggregator.

Once initialized, the Cartesi Machine runs a shell script that decompresses the contents of input drive~1 to access the per-household CSV files and their signatures.
The shell script then runs a Python~3 script that checks the signatures of each household CSV file using the OpenSSL utility and, if successful, inserts the consumption data into an SQLite database.
With the data loaded, the Python~3 script runs a variety of SQL queries that compute the amount owed by each household and, when done, stores the results into the output drive.

The electricity provider can perform the computation and post the final hash to the blockchain.
Each household can then perform the same computation and either dispute the results or pay what is justly owed.

For simplicity, our experiment was limited to the Cartesi Machine computation itself, since this is the only part relevant to our contribution: the new dispute resolution algorithm.
We used the Smart\textsuperscript{*} Microgrid dataset~\cite{BarkerEtAl2012} source for consumption data.
This dataset includes the average real power usage for 400 homes at one minute granularity for an entire day.
Given the availability of mainstream development tools, coding the entire application took less than a day of work.

We used a subset with 100 households and created 3 types of dishonest machine.
The first modifies the contents of the CSV file of a given household, right before it is read by the honest machine.
This fails the signature test.
The second modifies the data for a household \emph{right after} the signature test, thereby successfully changing the value owed by all households.
The final type of dishonest machine changes the result of a household after it has been stored in the output.
In total, there are 300 dishonest machines.\footnote{Designing and implementing these dishonest machines took considerably longer than implementing the honest application.}

\subsection{Results}

In order to simplify the setup, our experiments run the server and each party as separate processes on the same dual-socket 48-core machine (Intel Xeon Platinum 8259CL @ 2.50GHz, AWS r5n.24xlarge instance).

In both experiments, the single honest party was able to defend against hundreds of dishonest opponents.
Moreover, the load on the refereeing server was insignificant.
By design, the flurry of activity among dishonest parties has only logarithmic effect on the honest party as it plows through to victory.

In a test as short as the simple experiment from section~\ref{s:simple_ex}, the multi-stage tournament setup is unnecessary.

In the complex experiment, however, we chose a 4-stage tournament $\big(g_i\in \{24, 14, 7, 0\}\big)$.
Skipping $2^{24}$ steps between state-hash calculations reduces the overhead of generating the computation hash to a perfectly manageable~150\% (109s vs.\ 250s) using a single core.
A match takes about 8 minutes.
In our test machine, using a single core per process, the entire tournament (with 301 participants and 4 stages) completed in 128 minutes.
Since a tournament with 301 participants completes in 9 rounds, we estimate that, in a more realistic experimental setup where each party runs an independent core, the tournament would complete in approximately 72 minutes.

Using 8 cores to speedup hash computation brings the stage~1 overhead further down to 50\% (109s vs. 163s).
A match takes about 4 minutes.
A more realistic setup with 8 cores per party would bring the tournament time down to approximately 36 minutes.

\section{Conclusion}

Our new algorithm for dispute resolution gives any honest party an exponential advantage, both financial and computational, over any number of dishonest parties.
We hope it will serve as the foundation of a new generation of layer-2 projects that not only serve to address the scalability challenges in programmable blockchains, but that are at the same time more decentralized and equitable.
Moreover, we believe the added scalability, and the ensuing ability of running computations under a modern operating system, will bring forth a large number of mainstream developers that have so far avoided blockchain development due to its many difficulties.
All our code will be released in open-source.


%






\bibliography{intro,nxn}

\begin{thebibliography}{10}
\providecommand{\url}[1]{#1}
\csname url@samestyle\endcsname
\providecommand{\newblock}{\relax}
\providecommand{\bibinfo}[2]{#2}
\providecommand{\BIBentrySTDinterwordspacing}{\spaceskip=0pt\relax}
\providecommand{\BIBentryALTinterwordstretchfactor}{4}
\providecommand{\BIBentryALTinterwordspacing}{\spaceskip=\fontdimen2\font plus
\BIBentryALTinterwordstretchfactor\fontdimen3\font minus
  \fontdimen4\font\relax}
\providecommand{\BIBforeignlanguage}[2]{{%
\expandafter\ifx\csname l@#1\endcsname\relax
\typeout{** WARNING: IEEEtran.bst: No hyphenation pattern has been}%
\typeout{** loaded for the language `#1'. Using the pattern for}%
\typeout{** the default language instead.}%
\else
\language=\csname l@#1\endcsname
\fi
#2}}
\providecommand{\BIBdecl}{\relax}
\BIBdecl

\bibitem{pease1980reaching}
\BIBentryALTinterwordspacing
M.~Pease, R.~Shostak, and L.~Lamport, ``Reaching agreement in the presence of
  faults,'' \emph{J. ACM}, vol.~27, no.~2, pp. 228--234, apr 1980. [Online].
  Available: \url{https://doi.org/10.1145/322186.322188}
\BIBentrySTDinterwordspacing

\bibitem{lamport1982byzantine}
\BIBentryALTinterwordspacing
L.~Lamport, R.~Shostak, and M.~Pease, ``The byzantine generals problem,''
  \emph{ACM Trans. Program. Lang. Syst.}, vol.~4, no.~3, pp. 382--401, jul
  1982. [Online]. Available: \url{https://doi.org/10.1145/357172.357176}
\BIBentrySTDinterwordspacing

\bibitem{lamport1998part}
\BIBentryALTinterwordspacing
L.~Lamport, ``The part-time parliament,'' \emph{ACM Trans. Comput. Syst.},
  vol.~16, no.~2, pp. 133--169, may 1998. [Online]. Available:
  \url{https://doi.org/10.1145/279227.279229}
\BIBentrySTDinterwordspacing

\bibitem{ongaro2014search}
D.~Ongaro and J.~Ousterhout, ``In search of an understandable consensus
  algorithm,'' in \emph{Proceedings of the 2014 USENIX Conference on USENIX
  Annual Technical Conference}, ser. USENIX ATC'14.\hskip 1em plus 0.5em minus
  0.4em\relax USA: USENIX Association, 2014, pp. 305--320.

\bibitem{douceur2002sybil}
J.~R. Douceur, ``The sybil attack,'' in \emph{Peer-to-Peer Systems},
  P.~Druschel, F.~Kaashoek, and A.~Rowstron, Eds.\hskip 1em plus 0.5em minus
  0.4em\relax Berlin, Heidelberg: Springer Berlin Heidelberg, 2002, pp.
  251--260.

\bibitem{nakamoto2008bitcoin}
S.~Nakamoto, ``Bitcoin: A peer-to-peer electronic cash system,''
  \emph{Decentralized Business Review}, p. 21260, 2008.

\bibitem{garay2015bitcoin}
J.~Garay, A.~Kiayias, and N.~Leonardos, ``The bitcoin backbone protocol:
  Analysis and applications,'' in \emph{Advances in Cryptology - EUROCRYPT
  2015}, E.~Oswald and M.~Fischlin, Eds.\hskip 1em plus 0.5em minus 0.4em\relax
  Berlin, Heidelberg: Springer Berlin Heidelberg, 2015, pp. 281--310.

\bibitem{dwork1992pricing}
C.~Dwork and M.~Naor, ``Pricing via processing or combatting junk mail,'' in
  \emph{Advances in Cryptology --- CRYPTO' 92}, E.~F. Brickell, Ed.\hskip 1em
  plus 0.5em minus 0.4em\relax Berlin, Heidelberg: Springer Berlin Heidelberg,
  1993, pp. 139--147.

\bibitem{lamport1981password}
\BIBentryALTinterwordspacing
L.~Lamport, ``Password authentication with insecure communication,''
  \emph{Commun. ACM}, vol.~24, no.~11, pp. 770--772, nov 1981. [Online].
  Available: \url{https://doi.org/10.1145/358790.358797}
\BIBentrySTDinterwordspacing

\bibitem{wood2022ethereum}
\BIBentryALTinterwordspacing
G.~Wood, ``Ethereum: A secure decentralised generalised transaction ledger,''
  Yellowpaper, 2022, berlin version beacfbd -- 2022-10-24. [Online]. Available:
  \url{https://ethereum.github.io/yellowpaper/paper.pdf}
\BIBentrySTDinterwordspacing

\bibitem{Sharding-FAQs}
\BIBentryALTinterwordspacing
V.~Buterin, ``{On sharding blockchains},'' Wiki, 2018. [Online]. Available:
  \url{https://github.com/ethereum/wiki/wiki/Sharding-FAQs}
\BIBentrySTDinterwordspacing

\bibitem{Rollups}
\BIBentryALTinterwordspacing
------, ``{An Incomplete Guide to Rollups},'' Blog, 2021. [Online]. Available:
  \url{https://vitalik.ca/general/2021/01/05/rollup.html}
\BIBentrySTDinterwordspacing

\bibitem{rollup-roadmap}
\BIBentryALTinterwordspacing
------, ``A rollup-centric ethereum roadmap,'' Forum, 10 2020. [Online].
  Available:
  \url{https://ethereum-magicians.org/t/a-rollup-centric-ethereum-roadmap/4698}
\BIBentrySTDinterwordspacing

\bibitem{Feige1997}
U.~Feige and J.~Kilian, ``Making games short,'' in \emph{Proceedings of STOC},
  1997, pp. 506--516.

\bibitem{gennaro2010non}
R.~Gennaro, C.~Gentry, and B.~Parno, ``Non-interactive verifiable computing:
  Outsourcing computation to untrusted workers,'' in \emph{Annual Cryptology
  Conference}.\hskip 1em plus 0.5em minus 0.4em\relax Springer, 2010, pp.
  465--482.

\bibitem{goldwasser2015delegating}
S.~Goldwasser, Y.~T. Kalai, and G.~N. Rothblum, ``Delegating computation:
  interactive proofs for muggles,'' \emph{Journal of the ACM (JACM)}, vol.~62,
  no.~4, pp. 1--64, 2015.

\bibitem{CanettiRivaEtAl2013}
R.~Canetti, B.~Riva, and G.~N. Rothblum, ``Refereed delegation of
  computation,'' \emph{Information and Computation}, vol. 226, pp. 16--36,
  2013.

\bibitem{TeutschEtAl2017}
\BIBentryALTinterwordspacing
J.~Teutsch and C.~Reitwie{\ss}ner, ``A scalable verification solution for
  blockchains,'' Whitepaper, 2017. [Online]. Available:
  \url{https://people.cs.uchicago.edu/~teutsch/papers/truebit.pdf}
\BIBentrySTDinterwordspacing

\bibitem{Arbitrum2018}
H.~Kalodner, S.~Goldfeder, X.~Chen, S.~M. Weinberg, and E.~W. Felten,
  ``Arbitrum: Scalable, private smart contracts,'' in \emph{Proceedings of the
  27th USENIX Conference on Security Symposium}, 2018, pp. 1353--1370.

\bibitem{TeixeiraNehab2018}
\BIBentryALTinterwordspacing
A.~Teixeira and D.~Nehab, ``The {Core} of {Cartesi},'' Whitepaper, 2018.
  [Online]. Available: \url{https://cartesi.io/cartesi_whitepaper.pdf}
\BIBentrySTDinterwordspacing

\bibitem{Merkle1979}
R.~C. Merkle, ``Secrecy, authentication, and public key systems.'' Ph.D.
  dissertation, Stanford University, 1979.

\bibitem{Cartesi_Rollups}
\BIBentryALTinterwordspacing
E.~de~Moura, ``{Cartesi Rollups},'' Medium, 2021. [Online]. Available:
  \url{https://medium.com/cartesi/scalable-smart-contracts-on-ethereum-built-with-mainstream-software-stacks-8ad6f8f17997}
\BIBentrySTDinterwordspacing

\bibitem{micali2000computationally}
S.~Micali, ``Computationally sound proofs,'' \emph{SIAM Journal on Computing},
  vol.~30, no.~4, pp. 1253--1298, 2000.

\bibitem{ben2018scalable}
E.~Ben-Sasson, I.~Bentov, Y.~Horesh, and M.~Riabzev, ``Scalable, transparent,
  and post-quantum secure computational integrity.'' \emph{IACR Cryptol. ePrint
  Arch.}, vol. 2018, p.~46, 2018.

\bibitem{CanettiRivaEtAl2011}
R.~Canetti, B.~Riva, and G.~N. Rothblum, ``Practical delegation of computation
  using multiple servers,'' in \emph{Proceedings of the {ACM} Conference on
  Computer and Communications Security}, 2011, pp. 445--454.

\bibitem{bunz2018bulletproofs}
B.~B{\"u}nz, J.~Bootle, D.~Boneh, A.~Poelstra, P.~Wuille, and G.~Maxwell,
  ``Bulletproofs: Short proofs for confidential transactions and more,'' in
  \emph{2018 IEEE Symposium on Security and Privacy (SP)}.\hskip 1em plus 0.5em
  minus 0.4em\relax IEEE, 2018, pp. 315--334.

\bibitem{groth2018updatable}
J.~Groth, M.~Kohlweiss, M.~Maller, S.~Meiklejohn, and I.~Miers, ``Updatable and
  universal common reference strings with applications to zk-snarks,'' in
  \emph{Annual International Cryptology Conference}.\hskip 1em plus 0.5em minus
  0.4em\relax Springer, 2018, pp. 698--728.

\bibitem{Optimism_Rollups}
\BIBentryALTinterwordspacing
Optimism, ``{Understanding the Optimistic Ethereum Protocol},'' Corporate
  Website, 2021. [Online]. Available:
  \url{https://community.optimism.io/docs/protocol/protocol.html}
\BIBentrySTDinterwordspacing

\bibitem{Dworkin2015}
M.~J. Dworkin, ``{SHA-3} standard: Permutation-based hash and extendable-output
  functions,'' No. Federal Inf. Process. Stds.(NIST FIPS)-202., Tech. Rep.,
  2015.

\bibitem{WatermanAsanovic2017U}
A.~Waterman and K.~Asanovi\'{c}, \emph{The {RISC-V} Instruction Set
  Manual}.\hskip 1em plus 0.5em minus 0.4em\relax RISC-V Foundation, 2017, vol.
  I: User-Level ISA, version 2.2.

\bibitem{WatermanAsanovic2017P}
------, \emph{The {RISC-V} Instruction Set Manual}.\hskip 1em plus 0.5em minus
  0.4em\relax RISC-V Foundation, 2017, vol. II: Privileged Architecture,
  version 1.10.

\bibitem{Tilley2018}
\BIBentryALTinterwordspacing
A.~Tilley, ``Google, {Tesla} get behind challenge to {ARM} chip design,''
  \emph{The Information}, 2018. [Online]. Available:
  \url{https://www.theinformation.com/articles/google-tesla-get-behind-challenge-to-arm-chip-design}
\BIBentrySTDinterwordspacing

\bibitem{RISCV-tests}
\BIBentryALTinterwordspacing
``{RISC-V unit tests},'' GitHub repository, 2020. [Online]. Available:
  \url{https://github.com/riscv/riscv-tests}
\BIBentrySTDinterwordspacing

\bibitem{Brotli2016}
J.~Alakuijala and Z.~Szabadka, ``Brotli compressed data format,'' RFC 7932,
  2016.

\bibitem{BarkerEtAl2012}
S.~Barker, A.~Mishra, D.~Irwin, E.~Cecchet, P.~Shenoy, and J.~Albrecht,
  ``Smart*: An open data set and tools for enabling research in sustainable
  homes,'' in \emph{SustKDD: Workshop on Data Mining Applications in
  Sustainability}, 2012.

\bibitem{zhou2020solutions}
Q.~Zhou, H.~Huang, Z.~Zheng, and J.~Bian, ``Solutions to scalability of
  blockchain: A survey,'' \emph{Ieee Access}, vol.~8, pp. 16\,440--16\,455,
  2020.

\bibitem{Nakamoto2009}
\BIBentryALTinterwordspacing
S.~Nakamoto, ``Bitcoin: A peer-to-peer electronic cash system,'' Whitepaper,
  2009. [Online]. Available: \url{http://bitcoin.org/bitcoin.pdf}
\BIBentrySTDinterwordspacing

\bibitem{Wood2017}
\BIBentryALTinterwordspacing
G.~Wood, ``Ethereum: A secure decentralised generalised transaction ledger,''
  Yellowpaper, 2018, byzantium version e94ebda -- 2018-06-05. [Online].
  Available: \url{https://ethereum.github.io/yellowpaper/paper.pdf}
\BIBentrySTDinterwordspacing

\bibitem{Cairo2020}
S.~I. Ltd., ``Cairo language,''
  \url{https://github.com/starkware-libs/cairo-lang}, 2020.

\bibitem{Zinc2020}
M.~Labs, ``Zinc framework,'' \url{https://github.com/matter-labs/zinc}, 2020.

\bibitem{8962150}
Q.~{Zhou}, H.~{Huang}, Z.~{Zheng}, and J.~{Bian}, ``Solutions to scalability of
  blockchain: A survey,'' \emph{IEEE Access}, vol.~8, pp. 16\,440--16\,455,
  2020.

\bibitem{groth2016size}
J.~Groth, ``On the size of pairing-based non-interactive arguments,'' in
  \emph{Annual international conference on the theory and applications of
  cryptographic techniques}.\hskip 1em plus 0.5em minus 0.4em\relax Springer,
  2016, pp. 305--326.

\bibitem{bitansky2013succinct}
N.~Bitansky, A.~Chiesa, Y.~Ishai, O.~Paneth, and R.~Ostrovsky, ``Succinct
  non-interactive arguments via linear interactive proofs,'' in \emph{Theory of
  Cryptography Conference}.\hskip 1em plus 0.5em minus 0.4em\relax Springer,
  2013, pp. 315--333.

\bibitem{Rollups_Goldman}
\BIBentryALTinterwordspacing
D.~Goldman, ``{The State of Optimistic Rollup},'' Blog, 2020. [Online].
  Available:
  \url{https://medium.com/molochdao/the-state-of-optimistic-rollup-8ade537a2d0f}
\BIBentrySTDinterwordspacing

\end{thebibliography}


\nocite{*}

\end{document}